\documentclass[twocolumn,aps,prb,showpacs,floatfix]{revtex4}

\usepackage{epsf}
\usepackage[dvips]{graphicx}
\usepackage{bm}

\begin{document}

\title{
Effect of geometric and electronic structures \\
on the finite temperature behavior of Na$_{58}$, Na$_{57}$, and Na$_{55}$ clusters
}

\author {Mal-Soon Lee}   \email{mslee@unipune.ernet.in}
\author {D. G. Kanhere}  \email{kanhere@unipune.ernet.in}

\affiliation{
Centre for Modeling and Simulation, and Department of Physics,
University of Pune, Ganeshkhind, Pune 411 007, India.}

\date{\today}

\begin{abstract}
An analysis of the evolutionary trends in the ground state geometries of Na$_{55}$ 
to Na$_{62}$ reveals Na$_{58}$, an electronic closed--shell 
system, shows namely an electronically driven spherical shape leading to a
disordered but compact structure.
This structural change induces a strong {\it connectivity} of short bonds among the surface 
atoms as well as between core and surface atoms with inhomogeneous strength 
in the ground state geometry, which affects its finite--temperature behavior.
By employing {\it ab initio} density--functional molecular dynamics,
we show that this leads to two distinct features in specific heat curve compared to
that of Na$_{55}$: (1) The peak is shifted by about 100~K higher in temperature.    
(2) The transition region becomes much broader than Na$_{55}$. 
The inhomogeneous distribution of bond strengths results in a broad melting transition 
and the strongly connected network of short bonds leads to the highest 
melting temperature of 375~K reported among the sodium clusters.
Na$_{57}$, which has one electron less than Na$_{58}$, also possesses stronger short-- 
bond network compared with Na$_{55}$, resulting in higher melting temperature (350~K) 
than observed in Na$_{55}$.
Thus, we conclude that when a cluster has nearly closed shell structure not only 
geometrically but also electronically, it show a high melting temperature. 
Our calculations clearly bring out the size--sensitive nature of the
specific heat curve in sodium clusters.
\end{abstract}
\pacs{31.15.Ew,36.40.Cg,36.40.Ei,64.70.Dv,71.15.Mb}

\maketitle

\section{Introduction \label{intro}}

The finite--temperature behavior of sodium clusters has been attracted much 
attention since the pioneering experimental work by Haberland and 
co-workers.~\cite{H98_99, H02_05} 
These experiments reported melting temperatures of sodium clusters in the 
size range of 55 to 350 atoms. 
A great deal of effort has been spent to understand and 
explain the main puzzling feature, namely the irregular variation in the 
observed melting points.
Equally puzzling is the observation that Na$_{57}$ and Na$_{142}$, 
which is neither geometric shell--closing systems nor electronic shell--closing systems, 
show higher melting temperatures than vicinity systems such as Na$_{55}$ and Na$_{147}$ 
of geometric shell--closing ones or Na$_{138}$ of electronic shell--closing one, 
which is not explained so far.
Although it is expected that there is an intricate interplay between the electronic 
and geometric structures so far as finite temperature behavior is concerned, 
the consensus seems to be that the geometry dominates the melting characteristics 
and the effects of electronic structure are secondary.~\cite{H02_05, Aguado} 
This has been a surprise since the stability of these clusters is 
almost entirely dictated by the electronic structure.
{\bf According to our present calculations, it reveals that 
when a cluster with nearly electronic closed shell structure has  
nearly icosahedral ground state geometry, it shows a high melting temperature.}

Recently a very different aspect of the finite--temperature behavior has been
revealed in the experimental measurements of heat capacities of gallium and aluminum 
clusters,~\cite{jarrold-ga_al} namely the size--sensitive nature of their {\it shapes}. 
A definitive correlation between the shape of specific heat curve and the nature 
of the ground--state geometry in gallium cluster has also been
established by our group.~\cite{Ga}
Surprisingly, in spite of extensive work reported on the bench mark 
system of sodium clusters, there is no report of such a size sensitivity.

In this paper, we show the effect of electronic structure on the melting of sodium 
cluster by comparison to a geometric shell--closing system.
It is interesting to note that there are very few cases in sodium clusters 
where an electronic shell--closing system (N=8, 20, 40, 58, 138, ...) has 
a similar size as a geometric shell--closing one (N=55, 147, 309, ...).
For instance, a pair of clusters N=55 and N=58 differs by three atoms only.
Another example is a pair \{N=138 and 147\} having nine--atom difference.
The next such occurrence is at N=309 and N=338, which is not close enough. 
Therefore, the expected effects due to both the geometric and electronic magic numbers 
are likely to be seen prominently in \{55, 58\} range. 
The electronically closed--shell structure of Na$_{58}$ and nearly electronic shell closing 
system of Na$_{57}$ have slightly distorted icosahedral structures as their ground state geometries 
compared to the geometric shell--closing cluster of Na$_{55}$. 
Indeed, as we shall see, Na$_{58}$ and Na$_{57}$, have significantly 
different finite--temperature characteristics as compared to those of geometrically closed--shell 
cluster, Na$_{55}$.
Firstly, the shape of the specific heat curve is much broader than that of Na$_{55}$.
Secondly, it shows a peak (rather broad) at the temperature approximately 375~K for Na$_{58}$ 
and 350~K for Na$_{57}$.
We note that the observed melting temperature of Na$_{57}$ by Haberland and co--workers
~\cite{H98_99} is about 325~K which differs less than 8~$\%$ from our result. 
Specially, in case of Na$_{58}$ this is the first observation that a melting temperature of sodium 
cluster can be closed to that of the bulk.
Interestingly, it is observed that Na$_{58}$ shows high
abundance in mass spectra, indicating its high stability.~\cite{H98_99}
The paper is organized as follows: 
in Sec.\ \ref{comp}, we briefly mention computational details used. 
In Sec.\ \ref{results} we discuss our results, and summarize the
results in Sec.\ \ref{summ}.

\section{Computational Details \label{comp}}

We have carried out Born--Oppenheimer molecular--dynamics
simulations using ultrasoft pseudopotentials within the local--density
approximation~(LDA).~\cite{VASP}
The reliability of our calculation can be judged from the fact that 
our earlier calculations based on density--functional theory (DFT) have successfully reproduced
the melting temperature of Na$_n$($n$=55, 92, and 142).~\cite{Na55-142}
In order to get an insight into the evolutionary pattern of the geometries, 
we have obtained equilibrium structures for Na$_n$ ($n$=55 to 62).
Our thorough search of the lowest--energy structure is done by obtaining at least 
180 distinguishable equilibrium configurations for each of the clusters by using 
a combination of a basin--hopping algorithm and density--functional methods.
With obtained ground--state geometry of Na$_{58}$ and Na$_{57}$, we have carried out extensive 
{\it ab initio} constant--temperature simulations using a Nose--Hoover thermostat to
compare with geometrically closed shell system, Na$_{55}$.
These simulations have been carried out at 16 temperatures for Na$_{58}$ and at 12 temperatures 
for Na$_{57}$ in the range of 80K$\le$ T$\le$500K for the period of at least 210~ps (240~ps near 
the melting temperature), total simulation times of 2.7~ns and 3.5~ns, respectively. 
Our cell size used is 24$\times$24$\times$24~\AA$^3$, with the energy cutoff of 3.6~Ry whose 
reliability has been examined in our previous work.~\cite{Na55-142} 
With obtained ground--state geometry of Na$_{58}$ and Na$_{57}$, we have carried out extensive 
{\it ab initio} constant--temperature simulations using a Nose--Hoover thermostat to
compare with geometrically closed shell system, Na$_{55}$.
These simulations have been carried out at 12 temperatures for Na$_{57}$ and at 16 temperatures 
for Na$_{58}$ in the range of 80K$\le$ T$\le$500K for the period of at least 210~ps (240~ps near 
the melting temperature), total simulation times of 2.7~ns and 3.5~ns, respectively. 
Our cell size used is 24$\times$24$\times$24~\AA$^3$, with the energy cutoff of 3.6~Ry whose 
reliability has been examined in our previous work.~\cite{Na55-142} 
We use the multiple--histogram technique to calculate specific heat.  
More details of the methods can be found in Ref.~\onlinecite{SiSn}.
For the analysis, we have taken the last 165~ps data from each temperature, 
leaving at least first 45~ps for thermalization.

\section{Results and Discussion \label{results}}


\begin{figure}
 \epsfxsize 3.3in
 \epsffile{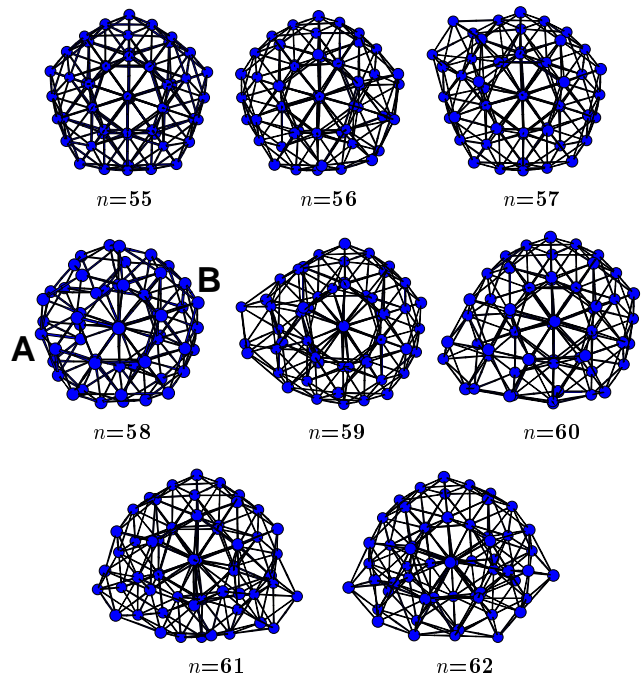} 
 \caption{
  The ground--state geometry of Na$_n$ ($n$=55-62).
  In the figure of Na$_{58}$, ``A" indicates the region where three extra atoms are 
  accommodated compared to Na$_{55}$, and ``B" indicates the region
  where the geometry of Na$_{55}$ is retained.}
\label{geometry}
\end{figure}
The lowest--energy structures of Na$_n$ ($n$=55--62) are shown 
in Fig.\ \ref{geometry}. 
There are some striking features evident in their evolutionary pattern.
The ground--state geometry (GS) of Na$_{55}$ is the well--known icosahedron~\cite{exp-na55}.
A single extra atom added to Na$_{55}$ is accommodated on the surface 
by a minor adjustment of surface atoms.
When the second atom is added, it is energetically more 
favorable to retain the icosahedral core with two atoms {\it capping} it.
Very interestingly, when three atoms are added ($n$=58), 
instead of the pattern of capping continuing, all three atoms are accommodated 
on the surface of icosahedron, making the structure 
again nearly spherical without changing the size of Na$_{55}$
significantly.
\begin{figure}
 \epsfxsize 3.3in
 \epsffile{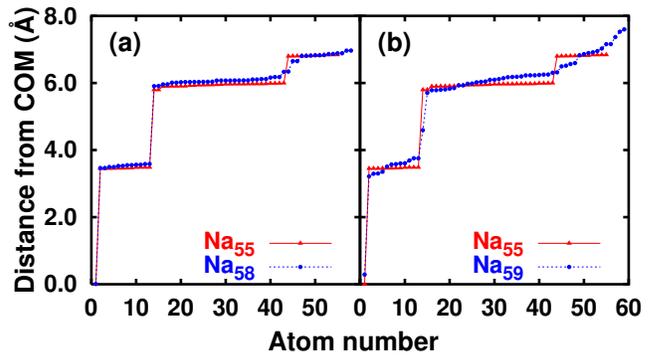}
 \caption{
  The distance from the center of mass of Na$_{58}$ and Na$_{59}$ in comparison 
  with that of Na$_{55}$. 
  In Na$_{58}$ the maximum distance is almost the same as seen in Na$_{55}$, 
  while that in Na$_{59}$ is changed significantly.}
\label{dist_from_COM}
\end{figure}
This can be verified by examining the distance from the center of mass (COM).
In Fig.\ \ref{dist_from_COM}, we show the distance from COM in Na$_{58}$ and 
Na$_{59}$ by comparison with those of Na$_{55}$.
It is clearly seen that in comparison with the maximum distance from COM in Na$_{55}$,
that in Na$_{58}$ is nearly remained the same, while in Na$_{59}$ it increases 
considerably. 
The pattern of growth from Na$_{59}$ to Na$_{62}$ changes back to the 
capping mode as can be seen in Fig.\ \ref{geometry}.
This peculiar shape transformation observed in Na$_{58}$ can be
examined by plotting deformation parameter $\varepsilon_{def}$.
$\varepsilon_{def}$ is defined as
$
\varepsilon_{def} = {2Q_1 / (Q_2+Q_3)}
$
, where $Q_1 \ge Q_2 \ge Q_3$ are eigenvalues of the quadrupole tensor
$
Q_{ij} = \sum_I R_{Ii}R_{Ij}
$
\noindent
with $R_{Ii}$ being {\it i}$^{th}$ coordinate of ion $I$ relative to the
center of mass of the cluster.
For a spherical shape ($Q_1 = Q_2 = Q_3$) $\varepsilon_{def}$ is 
1.0, while $\varepsilon_{def} > $ 1.0 indicates a deformation.
\begin{figure}
  \epsfxsize 2.8in
  \epsffile{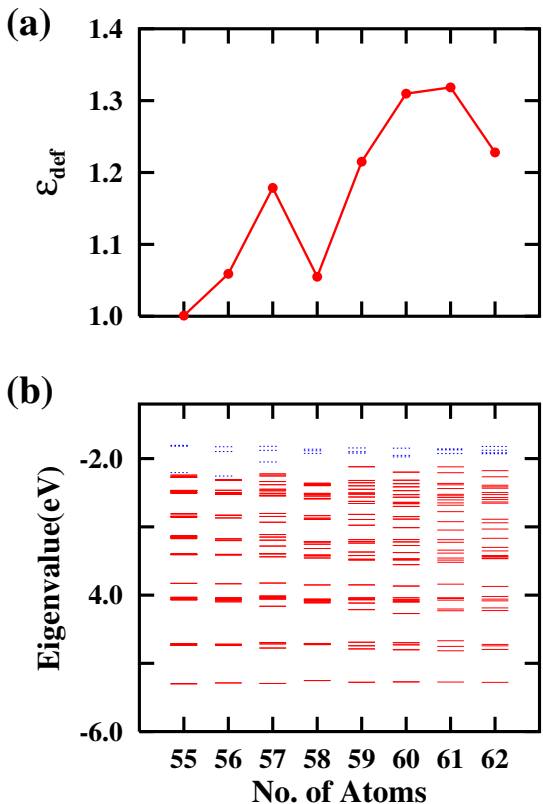}
  \caption{
   The comparison of (a) the deformation parameter, $\varepsilon_{def}$,  
   (b) the eigenvalue spectra of the ground--state geometries as a function of 
   cluster size.}
 \label{eps_def_evs}
\end{figure}
It can be seen in Fig.\ \ref{eps_def_evs}(a) that the addition of three atoms 
over Na$_{55}$ changes the shape to nearly spherical.
Interestingly, this difference is reflected in their eigenvalue 
spectrum as shown in Fig.\ \ref{eps_def_evs}(b).
A peculiarity of Na$_{58}$ is that the structure is not highly
symmetric, rather in the sense of amorphous.
However, it follows jellium--like pattern very closely in contrast 
to Na$_{57}$, Na$_{59}$ to Na$_{62}$ where additional states 
appear in the gaps due to their disordered structures.
Such an effect is absent in Na$_{58}$.
In addition an electronic shell--closing system of Na$_{58}$ shows 
the highest gap of the highest occupied molecular orbital (HOMO) 
and the lowest unoccupied molecular orbital (LUMO) among present studied
systems.
Thus, an electronically closed--shell nature of Na$_{58}$ results into a spherical 
charge density distribution which drives the geometry towards a spherical shape.
Our calculations bring out the fact that this change of shape induces
the GS geometry to be compact with a significant structural disorder in Na$_{58}$.
This is an example of an {\it electronically driven shape change}.  

\begin{figure}
\epsfxsize 3.2in
\epsffile{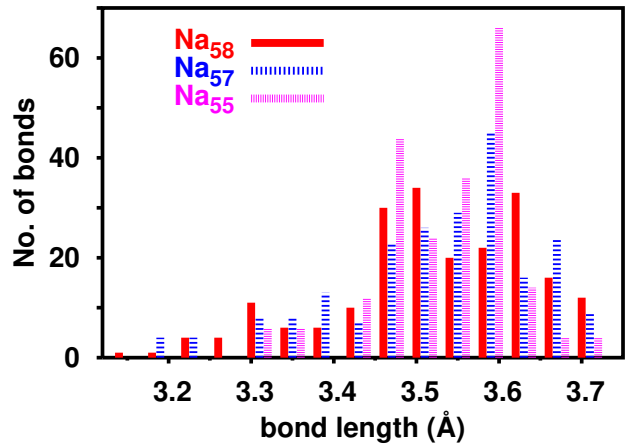}
 \caption{
  The histogram of bond lengths for the ground state geometry of Na$_{58}$, Na$_{57}$, and Na$_{55}$, 
  with less than that of the bulk (3.71\AA).}
\label{bondlength}
\end{figure}

\begin{figure}
 \epsfxsize 3.3in
 \epsffile{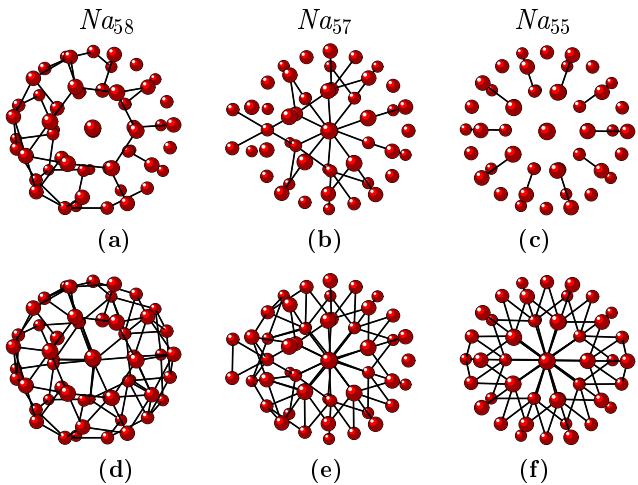}
 \caption{
  The short--bond connectivity with bond length less than 3.45\AA\ in
  (a) Na$_{58}$ (b) Na$_{57}$ (c) Na$_{55}$, showing inhomogeneous distribution of bond strength 
  in Na$_{58}$ and Na$_{57}$.
  The short--bond connectivity with bond length less than 3.55\AA\ in
  (d) Na$_{58}$ (e) Na$_{57}$ (f) Na$_{55}$, showing Na$_{58}$ has
  strongest connectivity through entire system among three clusters.\\ }
 \label{connectivity}
\end{figure}

We find that to understand the melting behavior of small cluster, which we shall see later,
it is important to study not only how many short bonds are there in the GS geometry of the cluster,
but also how these short bonds are connected each other (we call this ``{\it connectivity}").
Thus, we have examined the bond lengths in Na$_{58}$, Na$_{57}$, and Na$_{55}$ 
as well as their connectivity.
Fig.\ \ref{bondlength} shows the number of bonds having bond lengths less than 3.71~\AA, 
the bulk bond length. 
Na$_{58}$ has 21 bonds shorter than the shortest bond in Na$_{55}$, which are located 
in the region A, shown in Fig.\ \ref{geometry}, giving rise to an island of relatively strongly bonded 
atoms as compared to Na$_{55}$. 
In Na$_{57}$ there are 16 bonds which are shorter than the shortest bond in Na$_{55}$.
Fig.\ \ref{connectivity} shows how the nature of connectivity in these clusters with different bond lengths.
In Figs.\ \ref{connectivity}(a)--\ref{connectivity}(c), connectivity of short bonds are
shown with bond lengths less than 3.45~\AA, and Figs.\
\ref{connectivity}(d)--\ref{connectivity}(f) for less than 3.55~\AA.
It can be seen that Na$_{55}$ has fewer short bonds (12 bonds) than
Na$_{58}$ (41 bonds) and Na$_{57}$ (38 bonds).
The majority of these strong bonded atoms in Na$_{58}$ form a connected island 
in the region A shown in Fig.\ \ref{geometry}, while all short bonds
in Na$_{55}$ are radial.
It is also evident from Fig.\ \ref{connectivity}(d) 
that for bond lengths less than 3.55~\AA\  Na$_{58}$ shows not only that core atoms are 
strongly bonded to surface but that even surface atoms are bonded each other. 
In contrast to this, there is a connectivity only between core (first shell) and 
surface atoms but not among surface atoms in Na$_{55}$ (Fig.\ \ref{connectivity}(f)).
Na$_{57}$ shows stronger and inhomogeneous connectivity than those seen in Na$_{55}$ but weaker than 
those seen in Na$_{58}$ in both bond length regimes.
Thus, we conclude that to accommodate three extra atoms in Na$_{58}$
first--shell and surface distance is reduced, resulting in a strong network extending 
over entire cluster with inhomogeneous strength.
To compare the GS structures of Na$_{58}$ and Na$_{55}$, 
We also analyze the distribution of short bond lengths in Na$_{58}$ and Na$_{55}$ 
by comparing the distance from different shells.
It turns out that while Na$_{55}$ has shorter bond length between
center atom and first--shell, the average bond length 
between first--shell and outer--shell is shorter in Na$_{58}$.
In a recent work by Aguado and L{\'o}pez,~\cite{Aguado} they have attributed the higher 
melting temperature, seen in experiments of sodium clusters,~\cite{H98_99, H02_05} 
to the existence of shorter bonds between the surface and first--shell atoms.
Our calculations are consistent with this observation. 

\begin{figure}
 \epsfxsize 3.0in
 \epsffile{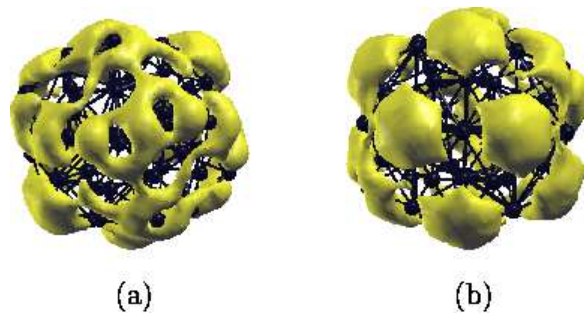}
 \caption{
  The isovalued surfaces of the electron localization function at 
  $\chi_{ELF}$=0.79 for Na$_{58}$ (a) in region A and (b) in region B.
  The similar pattern of (b) is also seen at $\chi_{ELF}$=0.76 in Na$_{55}$.} 
 \label{elf}
\end{figure}

We have examined the electron localization function (ELF).
The ELF~\cite{ELF} is defined as
$$
\chi_{\small ELF} = \left [ 1 + \left ( {D \over D_h} \right )^2 \right ]^{-1}
$$
\noindent
where
$$
 D  =  {1 \over 2} \sum_i \left | \nabla \psi_i \right |^2
     - {1 \over 8} {\left | \nabla \rho \right |^2  \over \rho}
$$
$$
 D_h = {3 \over 10} \left ( 3 \pi^2 \right )^{5/3} \rho^{5/3}
$$ 
\noindent 
Here $\rho \equiv \rho ({\bf r})$ is the valence electron 
density and $\psi_i$'s are the KS orbitals. The $\chi_{ELF}$ is 1.0 for 
perfect localized function and 0.5 for plain waves.  In Fig.\ \ref{elf}, 
we show the ELF isosurface of Na$_{58}$ for isovalue of $\chi_{ELF}$=0.79 
in two different regions. 
Figs.\ \ref{elf}(a) and \ref{elf}(b) depicts the ELF at 0.79 in regions A and
B, respectively. The contrast is evident. In region A which consists of 
atoms connected by shorter bonds, the ELF isosurface forms a connected 
network showing the existence of a strongly bonded region containing at 
least 20 atoms. In contrast, in region B such a network is absent. 
In fact, the region B gets connected at a lower isovalue of 0.74. 
The same analysis has been carried out for N$_{57}$ and Na$_{55}$.
Na$_{57}$ shows similar characteristics compared to those seen in Na$_{58}$.
A region near the capping atoms (left--hand side in Fig.\ \ref{connectivity}(b) 
starts connecting at $\chi_{ELF}$=0.78,
whereas a region away from them is connected at 0.74.
The connection of isosurface is established over the entire cluster of Na$_{55}$ 
at the isovalue of 0.72. 
Thus, our ELF analysis clearly brings out the existence 
of strongly bonded island of atoms and the inhomogeneous distribution of bond 
strength in Na$_{58}$ and Na$_{57}$.

Our study reveals two unique features in the ground state geometries of Na$_{58}$
and Na$_{57}$.
Firstly, the (nearly) electronic shell closing system induces shortening of
bonds, resulting in a strong connectivity of short bonds among the surface atoms 
as well as between first--shell and surface atoms compared to only
first--shell to core connectivity established in Na$_{55}$. 
Secondly, its ground--state geometry is considerably disordered,
resulting in inhomogeneous distribution of bond strength as compared to highly 
symmetric Na$_{55}$. 
Hear, disordering in these cases is related to the absence of spherical symmetry.
We argue that these two effects will be manifested in the specific heat differently. 
The existence of the well--connected short--bonds are expected to 
raise the melting temperature as compared to that of Na$_{55}$.
The effect of geometric disorder is to broaden the specific heat.

\begin{figure}
 \epsfxsize 3.0in
 \epsffile{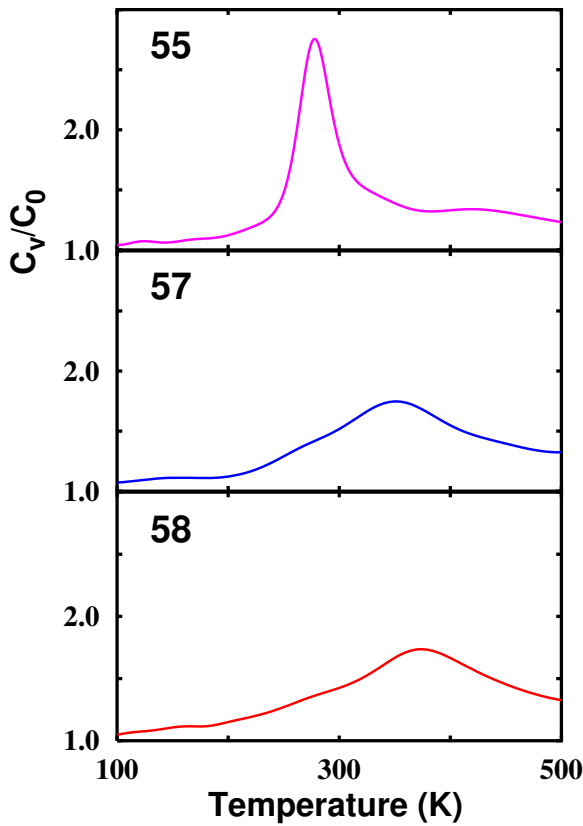}
 \caption{
  The normalized specific heat as a function of temperature.
  $C_0=(3N-9/2)k_B$ is the zero--temperature classical
  limit of the rotational plus vibrational canonical specific heat.}
\label{cv}
\end{figure}
We show the calculated specific--heat curves for Na$_{58}$ and Na$_{57}$ along 
with those of Na$_{55}$,~\cite{Na55-142} in Fig.\ \ref{cv}.
The most symmetric cluster Na$_{55}$ shows a sharp melting
transition at 280~K, while the highest melting temperature is
seen in Na$_{58}$.
Thus, the addition of two or three atoms changes the specific heat curve drastically.
Our calculations clearly demonstrate a strong correlation between the
nature of order or lack of order in the GS geometry and their finite
temperature behavior, and between the connectivity of short bonds and
its melting temperature.
This explanation can be extended to the finite temperature behavior in Na$_{50}$
(figure not shown).
Na$_{50}$ has a disordered GS geometry but does not possess a strongly connected 
network.~\cite{Na8-55} 
The specific--heat curve of Na$_{50}$, indeed, shows a very broad melting 
transition with low melting temperature (225~K).
The results of the density--functional simulations carried by Rytk{\"o}nen 
{\it et al.}~\cite{Manninen} as well as Lee {\it et al.}~\cite{Na8-55} 
are also consistent with the features that emerge out of our present 
calculation. 
An electronically closed--shell system Na$_{40}$ has a disturbed spherical 
GS geometry with a well--connected network of short bonds from core to surface 
atoms, which is stronger than that of Na$_{55}$.
Its specific heat is broader but melting point is higher than that of 
Na$_{55}$ (figure not shown).
Thus, it also brings out the size--sensitive nature of specific heat curve
in sodium clusters, not reported so far either experimentally or
computationally.
It may be noted that extreme size sensitivity of this kind has been
observed experimentally in Ga and Al clusters.~\cite{jarrold-ga_al}
We have also calculated the latent heats per tom by using caloric curves.
They are estimated as 0.02~eV/atom for Na$_{58}$, 0.016~eV for Na$_{57}$, 
and 0.014~eV/atom for Na$_{55}$.
The latent heats observed by experiment~\cite{H98_99} are 0.008~eV/atom 
for Na$_{57}$ and 0.015~eV/atom for Na$_{55}$. 
The deviation observed in Na$_{57}$ may be due to its very broad melting 
transition.

Due to the disordered nature of the GS geometry in Na$_{58}$, we expect
different atoms to move with various amplitudes at a given temperature.
To see this, we have calculated the Lindemann criteria, i.e., the 
root--mean--square bond length fluctuation ($\delta_{rms}$) for
individual atoms of Na$_{58}$, where $\delta_{rms}(i)$ for atom $i$ is defined as 
$$
\delta_{rms}(i) = { 1 \over N } \sum_j
               {\sqrt{\langle R_{ij}^{2} \rangle_{t}-
                      \langle R_{ij}     \rangle_{t}^{2}}
                  \over
                \langle R_{ij}\rangle_{t}}
$$
where $R_{ij}$ is the distance between $i$ and $j$ ions with {\it i}
$\ne${\it j}, $N=n-1$ with $n$ be number of atoms in the cluster, and 
$\langle$...$\rangle_t$ denotes a time average over the entire trajectory.
Indeed, $\delta_{rms}(i)$ in Na$_{58}$ has a broad range of values at a given
temperature showing a different response for each atom, 
while for Na$_{55}$ (highly symmetric structure) they are
collective, leading to a sharp transition region.
Further, it saturates at nearly 375~K in Na$_{58}$ compared to 280~K
in Na$_{55}$. 
Interestingly, the center atom in Na$_{58}$ (the bottom most line)
dose not develop its melting behavior till 300~K. 
\begin{figure}
  \epsfxsize 3.4in
  \epsffile{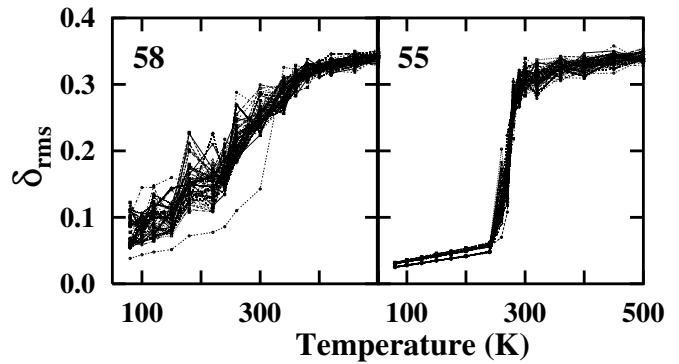}
  \caption{
   The comparison of the root--mean--square bond length 
   fluctuation ($\delta_{rms}$) of individual atoms for 
   Na$_{58}$ and Na$_{55}$.}
  \label{delta_atom}
\end{figure}
\begin{figure}
  \epsfxsize 3.3in
  \epsffile{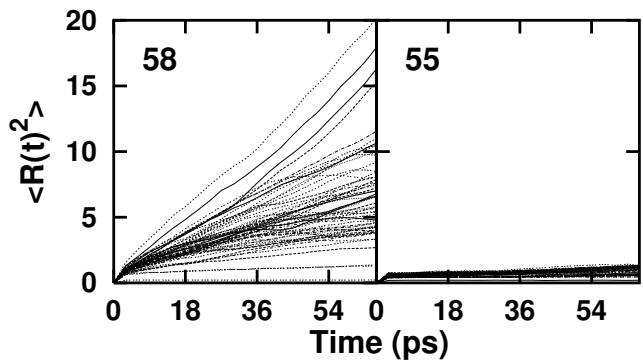}
  \caption{
   The comparison of the mean square displacement (MSD) of individual 
   atoms at 240~K for Na$_{58}$ and Na$_{55}$.}
  \label{atom_msq}
\end{figure}
The disordered nature of the geometry leading to inhomogeneous
distribution of bond strength results in distinctly different 
response of different atoms upon heating.   
We show the mean--square--displacement (MSD) of individual atoms
at 240~K in Fig.\ \ref{atom_msq}.
Evidently the atoms in these two clusters respond differently when
heated.
While Na$_{55}$ shows a clear solid--like behavior,
a significant number of atoms in Na$_{58}$ show diffusive motion.
We also observed the similar characteristics in Na$_{57}$, seen in
$\delta_{rms}$ and MSD of Na$_{58}$.

\begin{figure}
  \epsfxsize 3.0in
  \epsffile{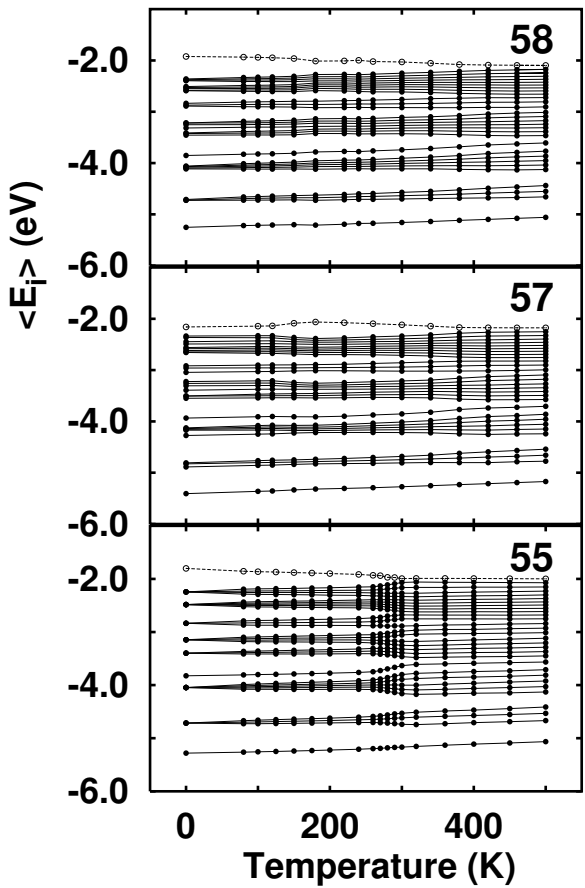}
  \caption{
   The average eigenvalues of Na$_{55}$, Na$_{57}$, and Na$_{58}$ as a function 
   of temperature.}
 \label{eigenvalue}
\end{figure}
\begin{figure}
  \epsfxsize 3.2in
  \epsffile{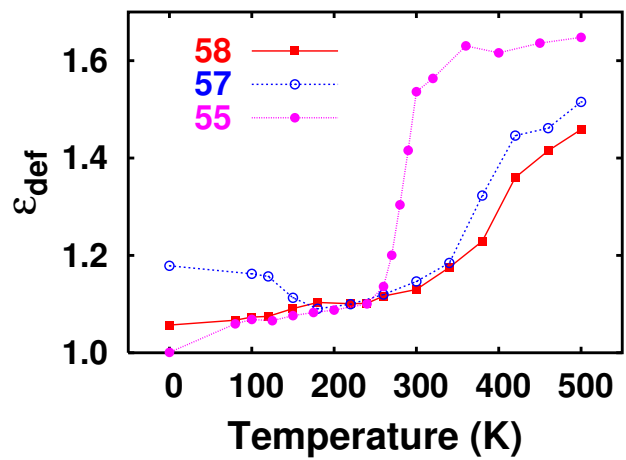}
  \caption{
   The average deformation parameter of Na$_{55}$, Na$_{57}$, and Na$_{58}$ as a function 
   of temperature.}
 \label{eps_def}
\end{figure}
As shown in Fig.\ \ref{eigenvalue} where we calculate 
the average eigenvalues carried over the entire simulation runs of Na$_{58}$, Na$_{57}$, 
and Na$_{55}$, we do not find that liquid--like and solid--like behavior of 
these clusters are electronically similar.
Instead, the HOMO--LUMO gaps in all the cases are closed.
According to the calculations of $\varepsilon_{def}$ as a
function of temperature, after melting their geometries become
elongated in these clusters.
This can explain the close of HOMO--LUMO gap.
Thus, the present extensive simulations clearly bring out the importance and
significance of the electronic structure.
Interestingly, the change of eigenvalue spectrum and $\varepsilon_{def}$ as a function of 
temperature in Na$_{58}$and Na$_{57}$ becomes similar after 180~K.
Even though the GS geometry of Na$_{57}$ is not spherical, its structure changes 
towards spherical shape upon heating.
Thus, when a cluster has both characteristics of (nearly) closed shell  
not only geometrically but also electronically, it has high melting temperature compared to
those having either geometric close shell or electronic closed shell.
This may explain the reason Na$_{142}$ has high melting temperature than Na$_{138}$, 
an electronic closed--shell cluster and Na$_{147}$, a geometric closed--shell one.  

\section{Summary \label{summ}} 

To summarize, the present work brings out the effect of the electronic
as well as the geometric structures on the melting of Na$_{58}$ and Na$_{57}$ clusters 
compared with that of Na$_{55}$.
The electronic shell--closing nature of Na$_{58}$ drives the GS geometry
to be spherical, which leads to a compact and disordered structure.
As consequence, in Na$_{58}$ the first--shell and surface as well as
atoms on the surface are well connected with short bonds.
This leads to the high melting temperature of nearly bulk melting temperature.
The disordered nature of system is responsible for rather broad
specific heat curve. 
Another example of strong connectivity is seen in the GS geometry of Na$_{57}$, nearly 
electronic shell closed cluster, which also shows high melting temperature than
geometrically closed shell system of Na$_{55}$. 
The strong correlation between connectivity of bonds and melting temperature is also 
seen in the case of Ga~\cite{Ga} and Al clusters.~\cite{Al}
Thus, we conclude that electronic structure affect its melting behavior strongly. 
We believe that it is possible to verify the prediction of present
work experimentally with the calorimetry method.

\section{Acknowledgments}

We acknowledge partial assistance from the Indo--French
Center for Promotion for Advance Research (IFCPAR).
We would like to thank Kavita Joshi and Sailaja Krishnamurty 
for a number of useful discussions.

\end{document}